\documentclass[preprint2]{aastex}

\shorttitle{Directionality of Flare-Accelerated Ions}
\shortauthors{Share \& Murphy}

\begin{document}


\title{Directionality of Solar Flare Accelerated Protons and $\alpha$ Particles from $\gamma$-Ray Line Measurements}


\author{Gerald H. Share and Ronald J. Murphy}
\affil{E.O. Hulburt Center for Space Research, Naval Research Laboratory, Washington, DC 20375 USA}
\author{J\"{u}rgen Kiener and Nicolas de S\'{e}r\'{e}ville}
\affil{CSNSM, In2P3-CNRS et Universit\'{e} Paris-Sud, B\^{a}timents 104 et 108, F-91405\\ Campus Orsay, France}


\begin{abstract}
The energies and widths of $\gamma$-ray lines emitted by ambient nuclei excited by flare-accelerated protons and $\alpha$ particles provide information on the ions directionality and spectra, and on the characteristics of the interaction region. We have measured the energies and widths of strong lines from de-excitations of $^{12}$C, $^{16}$O, and $^{20}$Ne in solar flares as a function of heliocentric angle.  The line energies from all three nuclei exhibit $\sim$1\% redshifts for flares at small heliocentric angles, but are not shifted near the limb.  The lines have widths of $\sim$3\% FWHM.  We compare the $^{12}$C line measurements for flares at five different heliocentric angles with calculations for  different interacting-particle distributions. A downward isotropic distribution (or one with a small upward component) provides a good fit to the line measurements.  An angular distribution derived for particles that undergo significant pitch angle scattering by MHD turbulence in coronal magnetic loops provides comparably good fits.   

\end{abstract}




\section{Introduction}

Gamma-ray lines emitted in solar flares provide information on the composition, spectrum, and angular distribution of accelerated ions that interact in the solar atmosphere \citep{rkl79}.  They also provide information on the composition and density of the ambient plasma.  

\citet{ram76} performed the first calculations of the spectral shape of emission lines due to Doppler shifts from recoil of excited nuclei after impact by protons and $\alpha$ particles.  They found that the $^{16}$O line at 6.13 MeV has a width of $\sim 100$ keV (FWHM) under bombardment by an isotropic distribution of protons.  This width increases to $\sim 160$ keV under bombardment by $\alpha$ particles.  They also estimated the shift of the line centroid, for a flare near the center of the solar disk, for both a downward isotropic distribution and a downward isotropic distribution limited to angles $\le 30\deg$ from the radial direction.  These estimates indicated red shifts of $\sim 25 $ and 40 keV for proton bombardment and $\sim 45$ and 60 keV for $\alpha$-particle bombardment, respectively. 

\citet{rkl79} performed detailed calculations of the shapes and shifts of gamma-ray lines from $^{12}$C and $^{16}$O that have become the basis of a Monte Carlo code fundamental to several subsequent publications.  \citet{murphy88} used updated cross sections in this code to calculate the solar $^7$Be, $^7$Li, and $^{12}$C line profiles for isotropic, fan beam, and downward beamed proton and $\alpha$-particle distributions having Bessel function energy spectra.  They also included contributions to the $^{12}$C line from spallation of $^{16}$O.  They described how these angular distributions can be distinguished from one another using gamma-ray line measurements made at different heliocentric angles, especially when made with a high-resolution spectrometer.  Based on an optical model with parameters determined by accelerator particle data, \citet{werntz90} developed a computer code that calculated the shapes of the $^{12}$C and $^{16}$O lines from inelastic collisions of protons.  They showed representative spectra for fan beams, downward beams, and isotropic interacting particle distributions for both thick and thin targets. \citet{lang91} expanded these line-shape calculations by including $\alpha$-particle interactions. 

\citet{hua87} provided the first estimate of the angular distribution of accelerated ions at the Sun from measurements of flare-produced neutrons at Earth.  They found that the data could only be fit by ions with either an isotropic distribution or a fan beam distribution produced by particles mirroring in a magnetic loop.   \citet{hua89} applied these transport calculations to the production of nuclear de-excitation $\gamma$-ray lines and calculated the time profiles of decaying phases of high-energy flares under the assumption that the energy release occurred over short time period relative to the emission.  Their comparisons with limited flare data indicated that transport effects of the particles in magnetic loops dominated the time profiles. \citet{murphy90} applied these transport calculations to determine the shapes of the $^7$Be and $^7$Li lines produced in $\alpha$-$^4$He fusion reactions.  They derived the angular distribution of interacting $\alpha$ particles for different amounts of pitch angle scattering caused by MHD turbulence in coronal magnetic loops.  The distribution approaches that of a fan beam when there is negligible pitch angle scattering. With increasing pitch-angle scattering a larger fraction of the $\alpha$ particles interact in the downward direction.  Because \citet{murphy90} were only able to compare their calculations with data from a limb flare, they were not able to constrain the amount of pitch-angle scattering encountered by the $\alpha$ particles.  \citet{share97} studied the $^7$Be and $^7$Li line profiles in 19 solar flares using the shapes calculated by \citet{murphy88}.  They found that a downward beam of $\alpha$ particles could be ruled out in two disk flares, and that isotropic and broad fan-beam distributions of interacting particles were both consistent with all the data.

Recently \citet{kiener01} have calculated the 4.439 MeV $^{12}$C line profiles in more detail than has been done previously.  They used new accelerator measurements and have combined the techniques developed by \citet{murphy88} and \citet{werntz90} to calculate the profiles over a broad range of interacting proton and $\alpha$-particle energies, including  contributions to the $^{12}$C line from spallation of $^{16}$O.  Their algorithm enables calculation of the line profiles for different interacting particle spectra, angular distributions, and $\alpha$/p abundances.    

\begin{figure}
\plottwo{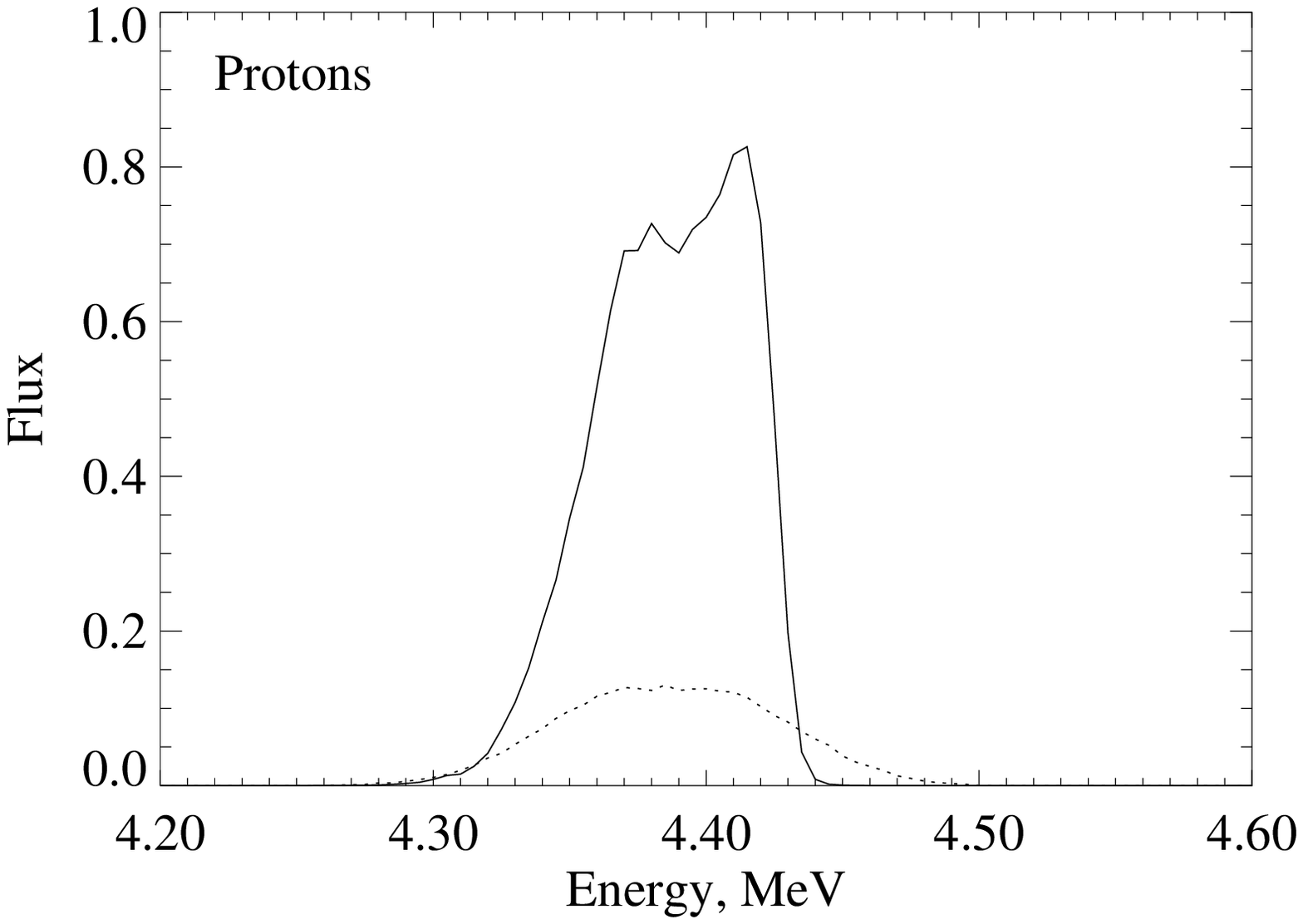}{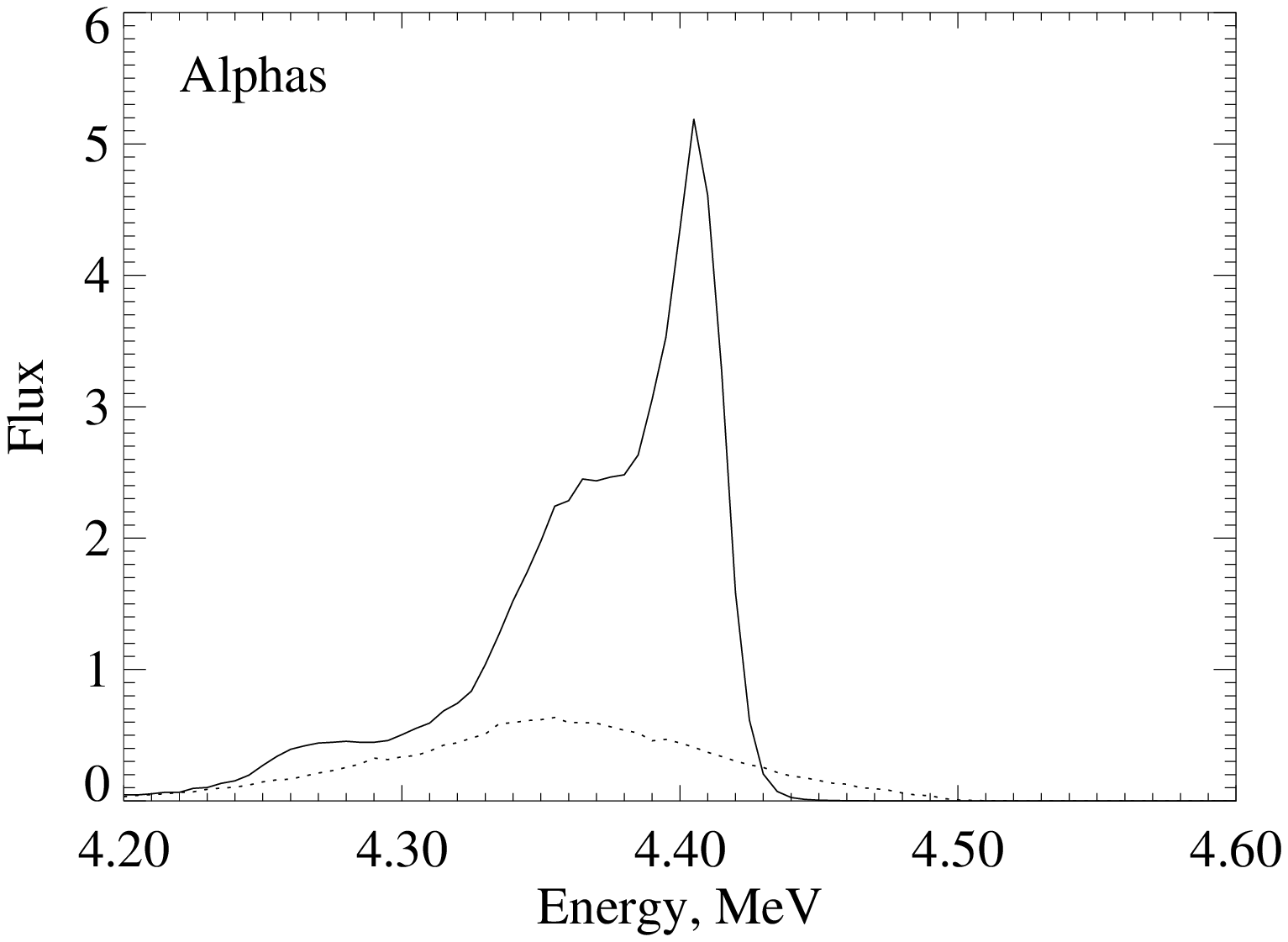}
\caption{Calculated $^{12}$C de-excitation line shapes observed at a heliocentric angle of 5$^{\circ}$ from impact of downward beams of protons  and $\alpha$-particles, following a power-law with index 4.0, on $^{12}$C (solid curves) and $^{16}$O (dotted curve) . \label{fig1}}
\end{figure}

In Figure~\ref{fig1} we plot calculated profiles that illustrate the various processes that contribute to the overall shape of the $^{12}$C line.  We assume that the proton and $\alpha$ particles follow power-law spectra with an index of 4.0 above $\sim$ 2 MeV/nucleon, are normalized to the same number of interacting particles, and are beamed ($5^{\circ}$ exponential fall-off) toward the Sun.  The calculations have been done for a flare at a heliocentric of 5$^{\circ}$.  In the left panel we plot the line profiles produced by the inelastic, $^{12}$C({\em p,p$\gamma$})$^{12}$C, and spallation, $^{16}$O({\em p,x$\gamma$})$^{12}$C, reactions.  The inelastic reaction dominates and produces a narrower line profile than the spallation reaction.  In the right panel we plot the inelastic $^{12}$C({\em $\alpha,\alpha\gamma$})$^{12}$C and 
spallation $^{16}$O({\em $\alpha,x\gamma$})$^{12}$C reactions.  The inelastic $\alpha$-particle reactions produce a more complex line profile extending to lower energy than that produced by the protons.  We have some concern about the $\alpha$-particle induced line shapes, however, because they are only based on optical model calculations that are most accurate for energies above $\sim$20 MeV/nucleon.  Laboratory measurements are an important priority. 

\begin{figure}
\plotone{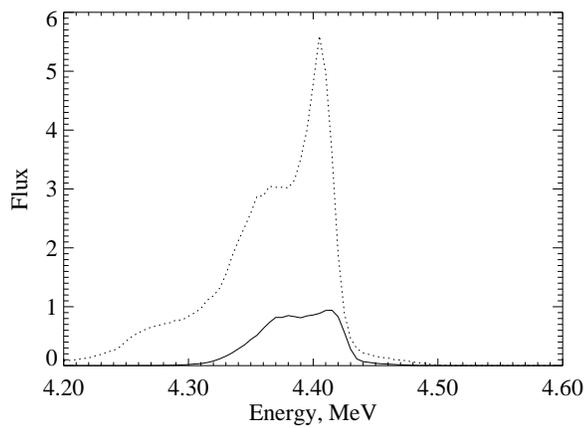}
\caption{Comparison of total $^{12}$C line shapes observed at a heliocentric angle of 5$^{\circ}$ from impact of downward beams of protons (solid curve) and $\alpha$-particles (dotted curves), following a power-law with index 4.0. \label{fig2}}
\end{figure}

We compare the total contributions to the $^{12}$C line profile from proton and $\alpha$-particle reactions in Figure~\ref{fig2}; the proton and $\alpha$-particle distributions are normalized to the same number of incident particles and once again follow a power-law spectrum with index 4.  The $\alpha$-particle induced reactions dominate the line shape for power-law accelerated-particle spectra and $\alpha$/proton ratios of $\sim$0.5 typically found in flares \citep{share97}.
 
In this paper we focus on measurements of the angular distributions of flare-accelerated protons and $\alpha$ particles that impact ambient $^{12}$C, $^{16}$O, and $^{20}$Ne through measurements of de-excitation lines from the recoiling nuclei.  We use spectral data from the {\it Solar Maximum Mission (SMM)} gamma-ray spectrometer experiment.  Although the spectrometer has only moderate spectral resolution it revealed $\sim$1\% red shifts for flares centered on the solar disk and $\sim$3\% intrinsic widths for all three lines.  We also compare the $^{12}$C line observations with calculations for different assumed angular distributions of interacting and protons and $\alpha$ particles.

\section{Observations}

The nuclear line measurements that we present were derived from 19 solar flares observed by {\it SMM} \citep{share95}.  The gamma-ray spectrometer \citep{forrest80} was exceptionally stable over its full $\sim$10 year mission.  It was also well calibrated.  The energy calibration was determined using laboratory sources and was confirmed by observations of  unshifted lines: the solar 0.511 MeV (positron annihilation) and 2.223 MeV (neutron capture) lines \citep{share00}, and the 4.444 MeV ($^{11}$B and $^{12}$C) and 6.13 MeV ($^{16}$O) lines from the Earth's atmosphere \citep{share01}.  Based on these measurements, we believe that line energies are known to $\la$3 keV.  The instrumental energy-resolution function was also determined using the laboratory calibrations and confirmed by the same flare and atmospheric line observations.  The instrumental resolution was 7.8\% (FWHM) at 0.511 MeV, 4.6\% at 2.223 MeV, 3.8\% at 4.44 MeV, and 3.4\% at 6.13 MeV.

\citet{share00} noted that the measured widths of de-excitation lines in the sum of the 19 flares were broader than predicted by theory for an isotropic distribution of particles.  Such broadening could occur if the interacting particle distribution is anisotropic (thus producing Doppler-shifted lines) because spectra from flares at various heliocentric angles were summed together.  We therefore summed spectra into 5 groups depending on the heliocentric angle of the flare.  The average angle and number of flares in each group are: 5$^{\circ}$, 1 (unfortunately, this was a relatively weak flare); 30$^{\circ}$, 3; 43$^{\circ}$, 5; 74$^{\circ}$, 5; 86$^{\circ}$, 5.  We plot count spectra in limited energy ranges around the 4.44 MeV $^{12}$C/$^{11}$B, 6.129 MeV $^{16}$O, and 1.634 MeV $^{20}$Ne de-excitation lines for the five heliocentric angles in Figures 3-5.  We fit the count spectra with a power law and a single Gaussian-shaped line with energy, width, and amplitude free to vary. Solid lines through the data represent the best fits.  For comparison, in Figure~\ref{fig6} we also plot the count spectra and fits for the thermal neutron capture line that is formed near rest and should appear at 2.223 MeV.  The vertical dotted lines are located at the rest energies of the nuclear lines.  The centroid energies of the de-excitation lines appear to shift to lower energies for flares close to the center of the solar disk.

We summarize the results on line shifts in Figure~\ref{fig7} where the best fit line energies and uncertainties for the three de-excitation lines and the neutron capture line are plotted as a function of the cosine of heliocentric angle.  The line energies and errors were derived by passing an incident photon spectrum, containing a power law and 22 lines from 0.3 to 8 MeV \citep{share01} through the instrument response function, and mapping $\chi ^2$; this yielded the same results as we obtained by simply fitting the counts spectrum with a power law and single Gaussian over a restricted energy range (see Figures 3 - 5).  

The $^{12}$C, $^{16}$O, and $^{20}$Ne de-excitation lines appear to be shifted to lower energies by $\sim$1\% near Sun center.   In contrast, the neutron capture line does not exhibit a significant redshift ($\la 0.1$\%) at Sun center.  The higher fitted energies ($\sim 0.3$\%) near the limb may be due to a contribution from other lines, e.g. $^{32}$S and $^{14}$N that become significant with limb darkening of the 2.223 MeV capture line.

The statistical significance of the line data in Figures 3-5 was also adequate for us to independently fit for the Gaussian widths of the $^{12}$C, $^{16}$O, and $^{20}$Ne de-excitation lines for some of the heliocentric angles.  We once again fit the count spectra and obtained the line widths and uncertainties using $\chi ^2$ mapping.  The line widths derived from these fits with smoothly varying $\chi ^2$ maps are plotted in Figure~\ref{fig8} for the three lines.  There is no significant variation in the widths as a function of heliocentric angle.  The weighted means of the line widths (FWHM) are 155 $\pm$ 19 keV, 167 $\pm$ 26 keV, and 40 $\pm$ 10 keV for the $^{12}$C, $^{16}$O, and $^{20}$Ne lines, respectively.  These are $\sim$3\% of the line energies.  Earlier calculations (e.g. \citet{murphy85}) suggested widths of $\sim$2\% and 2.6\% for the carbon line produced by direct excitation and spallation, respectively, from proton interactions.  The widths are expected to be even broader if accelerated $\alpha$ particles play a significant role in line production \citep{ram76,lang91}.

\section{Comparisons with Models}

We have calculated the shape of the $^{12}$C de-excitation line for different interacting particle distributions based on the recent work of \citet{kiener01}.  Line shapes were calculated for the following angular distributions: downward beam (5$^{\circ}$ exponential fall-off); fan beam following sin$^{6} \theta$ where $\theta$ is the angle between the particle velocity and sun center \citep{murphy88}; downward isotropic; and upward isotropic.  These calculations were performed for flares at the heliocentric angles of the five grouped {\em SMM} observations.  From a comparison of the fluxes in the 1.63 MeV ($^{20}$Ne) and 6.13 MeV ($^{16}$O) lines for the flares grouped by heliocentric angle, we obtained power-law spectral indices from $\sim$3.9 to 4.5  for accelerated particles having impulsive-flare abundances and an ambient Ne/O ratio of 0.25 \citep{ram96}.  In Figure~\ref{fig9} we plot the calculated line profiles for a downward isotropic angular distribution of accelerated particles following a power-law with spectral index 4.0 at the five heliocentric angles.   The solid and dotted curves show the shapes for an assumed accelerated $\alpha$/p ratio of 0.5 and 0.1, respectively.  It is clear that the $\alpha$ particles play a dominant role in determining the line shapes when the spectra are this steep.  We note that the line width produced by $\alpha$ particles is significantly broader than that produced by protons.  

\begin{figure}
\plotone{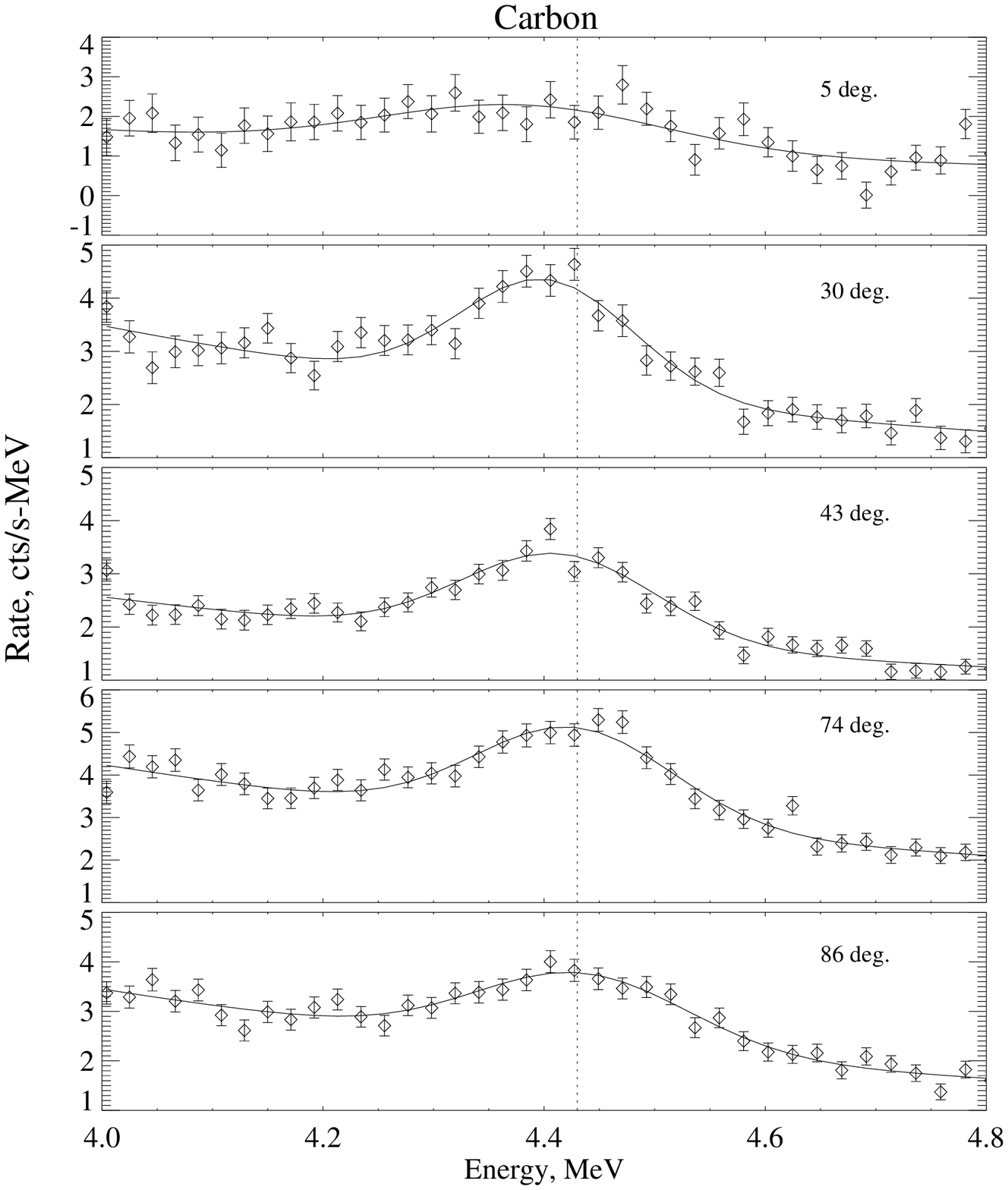}
\caption{Gaussian line fits to the $^{12}$C de-excitation line observed in flare spectra at different heliocentric angles.  The dotted line shows the rest energy of the de-excitation. \label{fig3}}
\end{figure}

We then folded the calculated line profiles for the particle distributions and a power-law continuum through the {\em SMM}/GRS instrument response function and fit the count data from 3.8 to 4.8 MeV.  The goodness of fit, measured using the $\chi ^2$ statistic, is the probability that a fit to a random distribution of numbers about a mean would give a higher value of $\chi ^2$ for the same number of degrees of freedom.  We found that the derived probabilities were not very sensitive to either the index of the power law or the $\alpha$/p ratio.  They varied by $<$30\% for indices ranging from 4 to 5 and $\alpha$/p ratios from 0.1 to 0.5. We summarize the results of these fits for the flare spectra accumulated at the 5 heliocentric angles in Table~\ref{tbl-1} for accelerated power-law particle spectra having an index of $\sim$4.2 and $\alpha$/p ratio of $\sim$0.3 (typical of flare observations).  Results from the Gaussian fits shown in Figure~\ref{fig3}, where the line width and energy were free to vary, are listed for comparison.  A downward beam of accelerated particles is ruled out by the fits.  Both isotropic and broad fan beam distributions also fail to provide acceptable fits to the data.  In contrast, a predominately downward distribution of isotropic interacting particles provides acceptable fits, comparable to that obtained by the Gaussian fits.  There is a small improvement in some of the fits when an upward isotropic distribution with $\sim$20\% of the amplitude of the downward isotropic distribution is added (see probabilities listed in the parentheses) in column 6.    

The angular distributions that we have used in the fits above are approximations to distributions that are derived under more realistic conditions.  \citet{hua89} and \citet{murphy90} calculated the angular distributions of interacting particles after transport in magnetic loops where the field is constant in the corona and converges from the chromosphere to the photosphere.  This calculation takes into account both magnetic mirroring and pitch angle scattering. \citet{hua89} define a parameter $\lambda$ as the scattering mean free path divided by the half length of the coronal segment of the magnetic loop.  They show that pitch angle scattering saturates for $\lambda$ = 25; particles are not scattered for $\lambda \rightarrow \infty$.  In the left panel of Figure~\ref{fig10} we plot the angular distributions of selected interacting particle distributions.  They are plotted versus cos $\theta$, where $\theta$ is the angle relative to the solar radius.  We see that the downward isotropic  distribution is a step function in this representation and the fan beam (sin$^{6} \theta$) shows a smoothly falling symmetric distribution about the parallel to the solar surface.  The weak pitch-angle scattering distribution, $\lambda$ = 300, begins to approach the fan beam but still has a downward component. The strong pitch-angle scattering distribution, $\lambda$ = 30, shows a broad downward distribution.

In the right panel of Figure~\ref{fig10} we plot the calculated $^{12}$C line profiles for the four angular distributions plotted in the left panel.  These calculations were done for a flare at a heliocentric angle of 30$^{\circ}$, an accelerated $\alpha$/p ratio of 0.5, and power law spectral index of 4.0. The downward isotropic and $\lambda$ = 30 distributions are similar.  Their shapes are indistinguishable for the moderate resolution {\em SMM} detectors as reflected in the fits listed in Table~\ref{tbl-1}.  The weak pitch angle scattering ($\lambda$ = 300) distribution has a blue wing that results in a factor of two poorer fit at 30$^{\circ}$.  The fan beam distribution is significantly different and is clearly incompatible with the {\em SMM} line observations. 

\section{Summary and Discussion}

We have measured the line energies and widths of $\gamma$-ray lines produced by excitation of ambient $^{12}$C, $^{16}$O, and $^{20}$Ne by solar flare protons and $\alpha$ particles.  The $\gamma$-ray spectra from nineteen solar flares observed by the {\em Solar Maximum Mission} satellite were grouped by heliocentric angle and were initially fit using Gaussian shaped lines.  The goodness of fit was characterized using the $\chi ^2$ statistic and the results are listed in Table~\ref{tbl-1}.  The Gaussians adequately represented the line shapes obtained with the moderate resolution NaI detectors.  The energies of each of the lines were consistent with their rest values for flares observed at the solar limb.  The energies of each of the lines decreased by $\sim$1\% for flares observed near the center of the solar disk.  All three lines exhibit Doppler broadening of $\sim$3\% FWHM of their rest energies ($^{12}$C, 155 $\pm$ 19 keV; $^{16}$O, 167 $\pm$ 26 keV; $^{20}$Ne, 40 $\pm$ 10 keV).  These results are suggestive of a broad angular distribution of particles primarily interacting in the downward direction.

The line shapes are determined by the location of the flare, the spectra, angular distribution, and composition of the interacting particles, and the intrinsic shapes determined by nuclear physics  \citep{rkl79,werntz90,kiener01}.  We therefore fit the same {\em SMM} spectral data obtained at the five heliocentric angle with line shapes derived for downward beam, fan beam, downward isotropic, and isotropic interacting particle distributions.  The results of these fits are also given in Table~\ref{tbl-1}.  A downward beam is clearly ruled out, while a broad fan beam and an isotropic distribution also appear to be inconsistent with the spectral data.  Only a downward isotropic distribution gives acceptable fits, comparable to those derived for Gaussian shaped lines.  We note that there is some improvement in the fits when we assume a distribution containing downward and upward isotropic components in the ratio of 5:1.  Thus the data are consistent with a very broad angular distribution of particles that primarily interact in the solar atmosphere in the downward direction.

It is important to relate these results on the distribution of interacting particles to their initial acceleration and transport in the solar magnetic loops.   We applied the transport and pitch angle scattering model developed by \citet{hua89} and \citet{murphy90} to calculate angular distributions of interacting particles.  We then folded these distributions into the algorithm that calculates the line shapes \citep{kiener01}.  From Table~\ref{tbl-1} we see that a scattering parameter, $\lambda$, of 30 provides comparably good line fits to those obtained for a Gaussian and for the downward dominated isotropic distribution.  Increasing the scattering parameter (decreasing the amount of pitch angle scattering) by a factor of 10 has only a relatively small effect on the goodness of the fits.  It is only when pitch angle scattering completely vanishes, and the particle distribution more closely follows that of a symmetric fan beam parallel to the photosphere, that the model no longer is consistent with the data.  This demonstrates that MHD turbulence is significant in the coronal magnetic loops.

As we have discussed earlier, $\alpha$ particles play a significant role in determining the $\gamma$-ray line shapes for power-law spectra with indices $\ga$4.0, especially if the accelerated $\alpha$/p ratio is $>$0.1. In our study of the $^{12}$C line shape we noted a small, but not significant, improvement in our fits when we increased the accelerated $\alpha$/p ratio to 0.5.  \citet{share97} presented evidence that the average accelerated $\alpha$/p ratio is in fact closer to $\sim$0.5 for photospheric $^4$He abundances in the interaction region \citep{man97}.  

\citet{share97} also studied the line shapes of the $\alpha$-$^4$He fusion lines in 19 individual flares and concluded that a downward beam of particles was inconsistent with the data, but that isotropic or fan beam distributions could fit the data.  We have repeated that study using the spectra summed by heliocentric angle.  We find that the spectral resolution of the {\em SMM} NaI detector is not adequate for us to unambiguously distinguish between isotropic, downward isotropic, and fan beam accelerated particle distributions using the $\alpha$-$^4$He fusion lines.  We expect that the high-resolution spectrometer on board the recently launched {\em HESSI} satellite \citep{lin00} will be able to distinguish between these distributions, and will determine if the same angular distribution of accelerated particles is responsible for producing both the $\alpha$-fusion  and de-excitation lines.  This same spectrometer will also make more definitive measurements of the amount MHD pitch angle scattering in the coronal magnetic loops.

\acknowledgments

We are grateful to  Benz Kozlovsky and Carl Werntz for discussions related to calculations of the $^{12}$C line shapes.  We also wish to thank Xin-Min Hua for his assistance with calculations of particle transport in magnetic fields and Vincent Tatischeff for discussions and careful reading of the manuscript.  This work was partially supported by NASA DPR 18995.

\clearpage

\clearpage

\clearpage

\clearpage 

\begin{figure}
\plotone{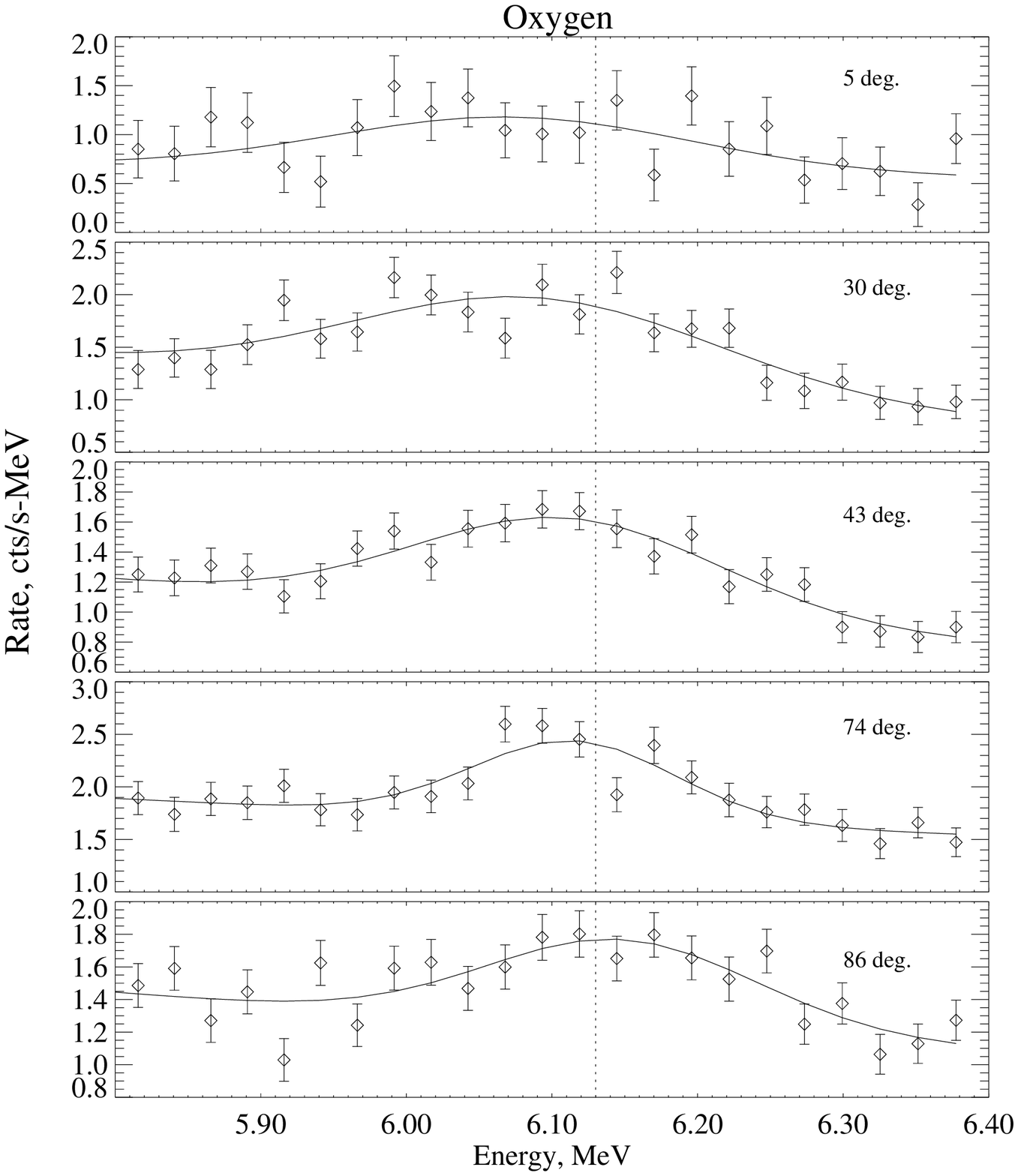}
\caption{Gaussian line fits to the $^{16}$O de-excitation line observed in flare spectra at different heliocentric angles.  The dotted line shows the rest energy of the de-excitation. \label{fig4}}
\end{figure}

\clearpage

\begin{figure}
\plotone{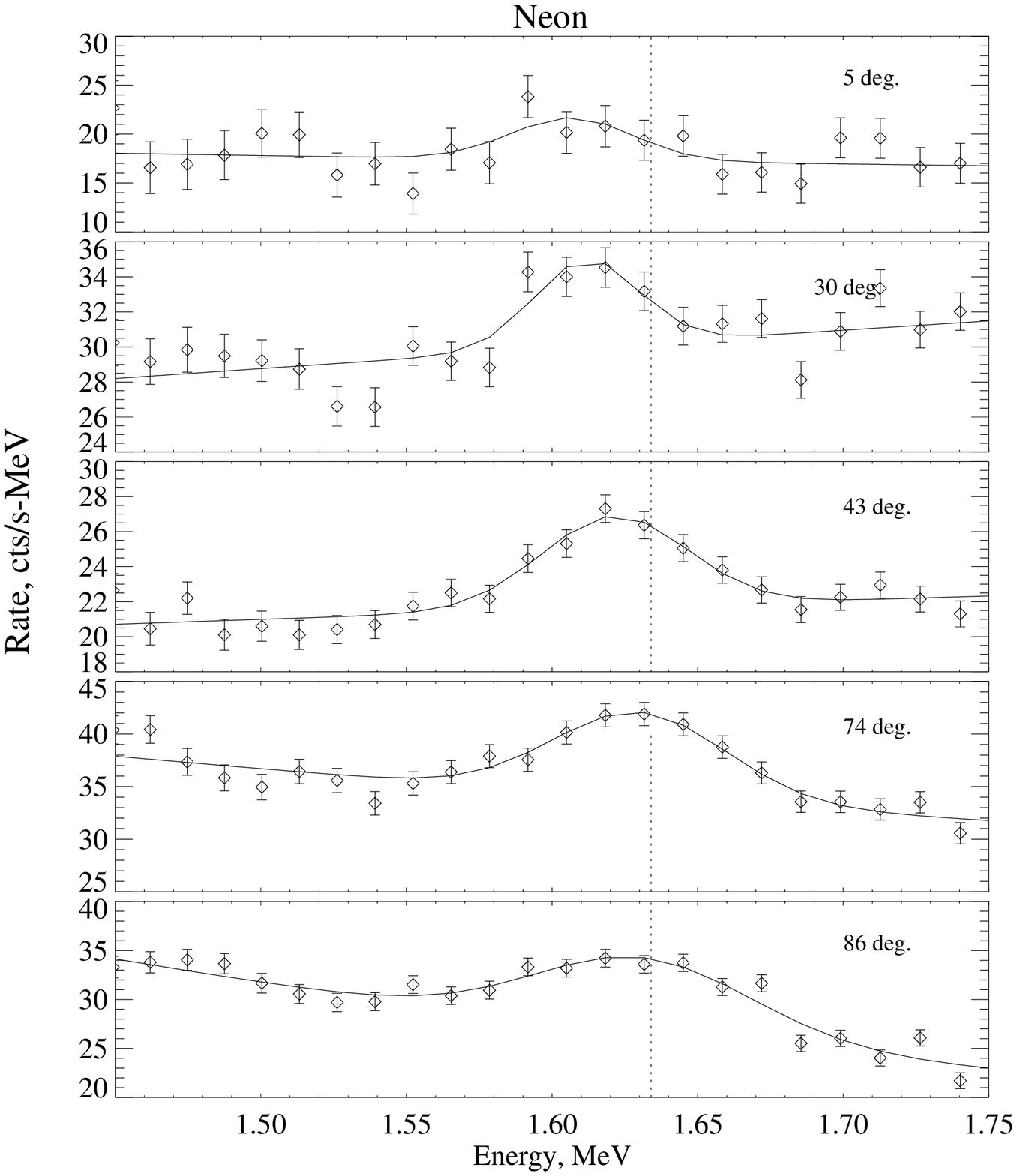}
\caption{Gaussian line fits to the $^{20}$Ne de-excitation line observed in flare spectra at different heliocentric angles.  The dotted line shows the rest energy of the de-excitation. \label{fig5}}
\end{figure}

\clearpage

\begin{figure}
\plotone{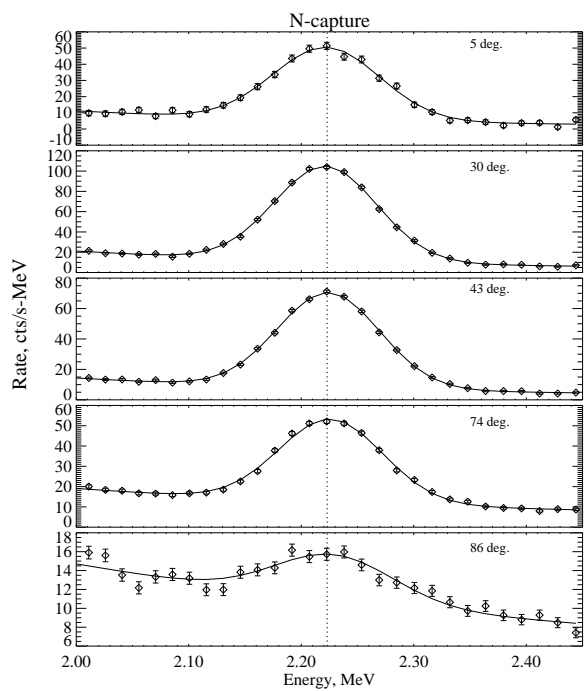}
\caption{Gaussian line fits to the neutron capture line observed in flare spectra at different heliocentric angles.  The dotted line shows the rest energy.} \label{fig6}
\end{figure}
\clearpage

\begin{figure}
\plotone{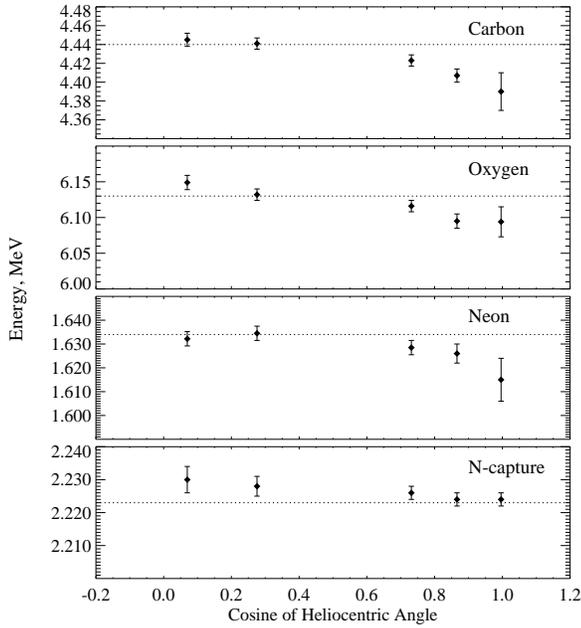}
\caption{Best fitting line energies for the $^{12}$C, $^{16}$O, $^{20}$Ne, and neutron capture lines as a function of cosine of heliocentric angle. } \label{fig7}
\end{figure}

\begin{figure}
\plotone{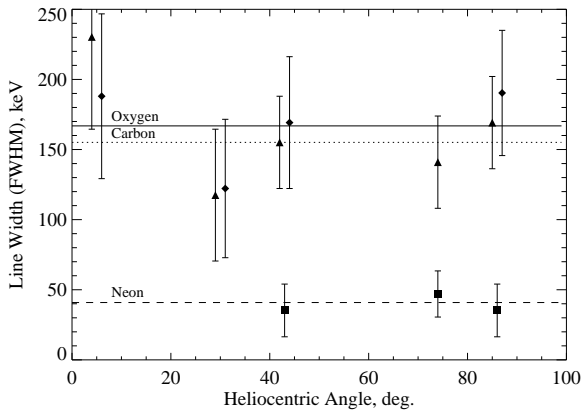}
\caption{Best fitting line widths (keV, FWHM) for the $^{12}$C ($\triangle$), $^{16}$O ($\Diamond$), and $^{20}$Ne ($\Box$) $\gamma$-ray lines vs heliocentric angle.  The dotted lines are the weighted averages of the widths.} \label{fig8}
\end{figure}

\begin{figure}
\plotone{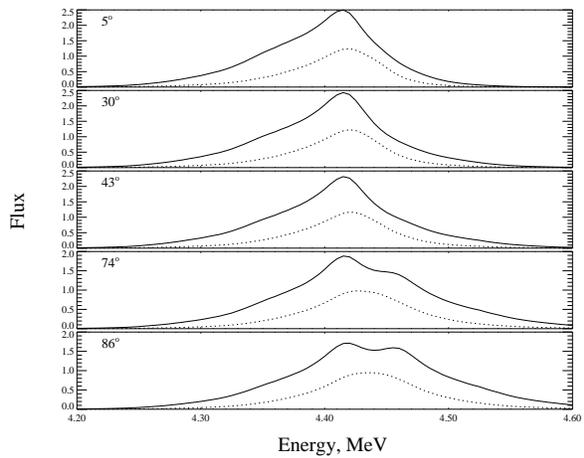}
\caption{Calculated $^{12}$C line profiles for a downward isotropic distribution of accelerated particles following a power law spectrum with index 4.0 for different heliocentric angles.  Accelerated $\alpha$/p = 0.5 (solid curve); accelerated $\alpha$/p = 0.1 (dotted curve)} \label{fig9}
\end{figure}
\clearpage

\begin{figure}
\plottwo{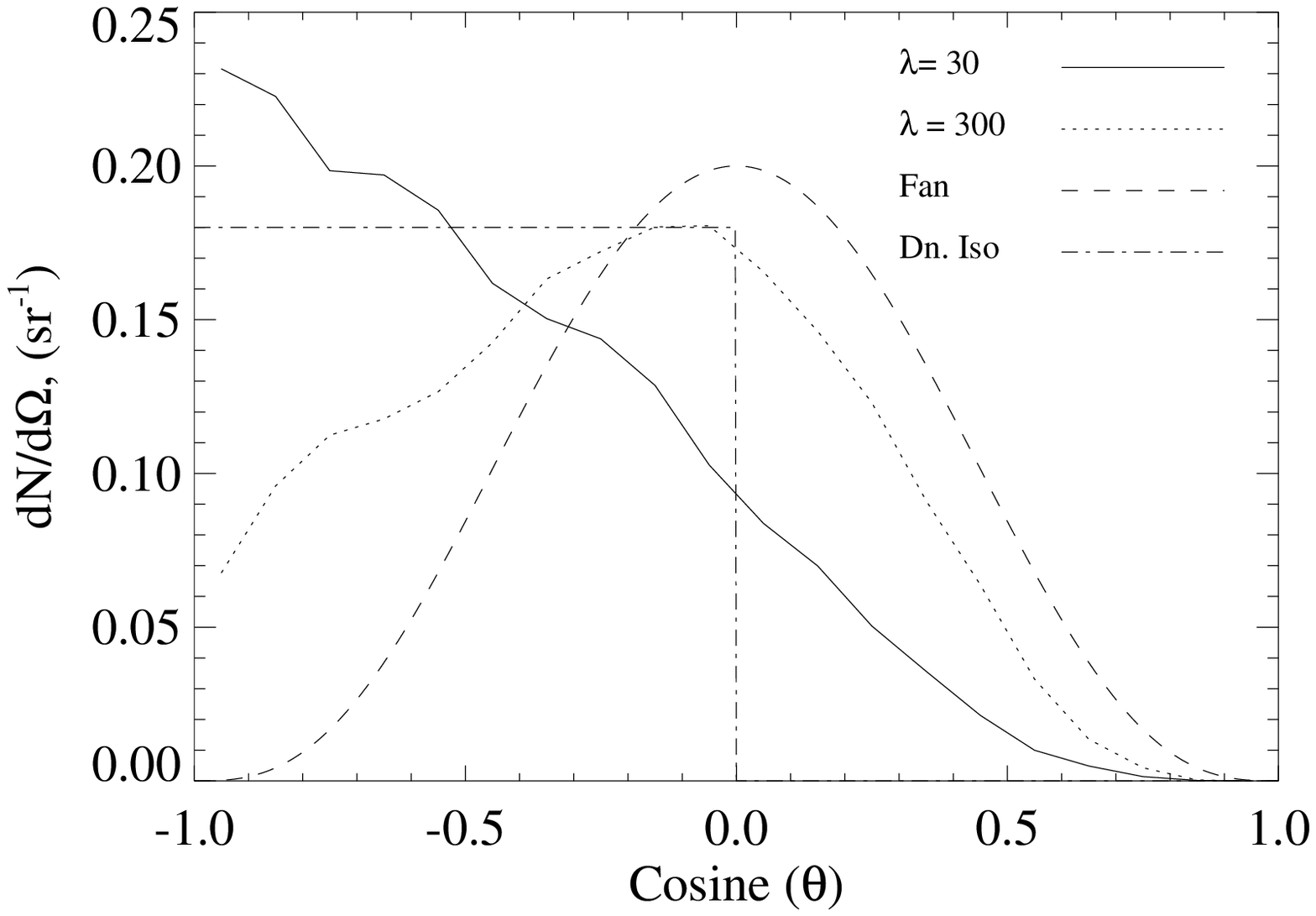}{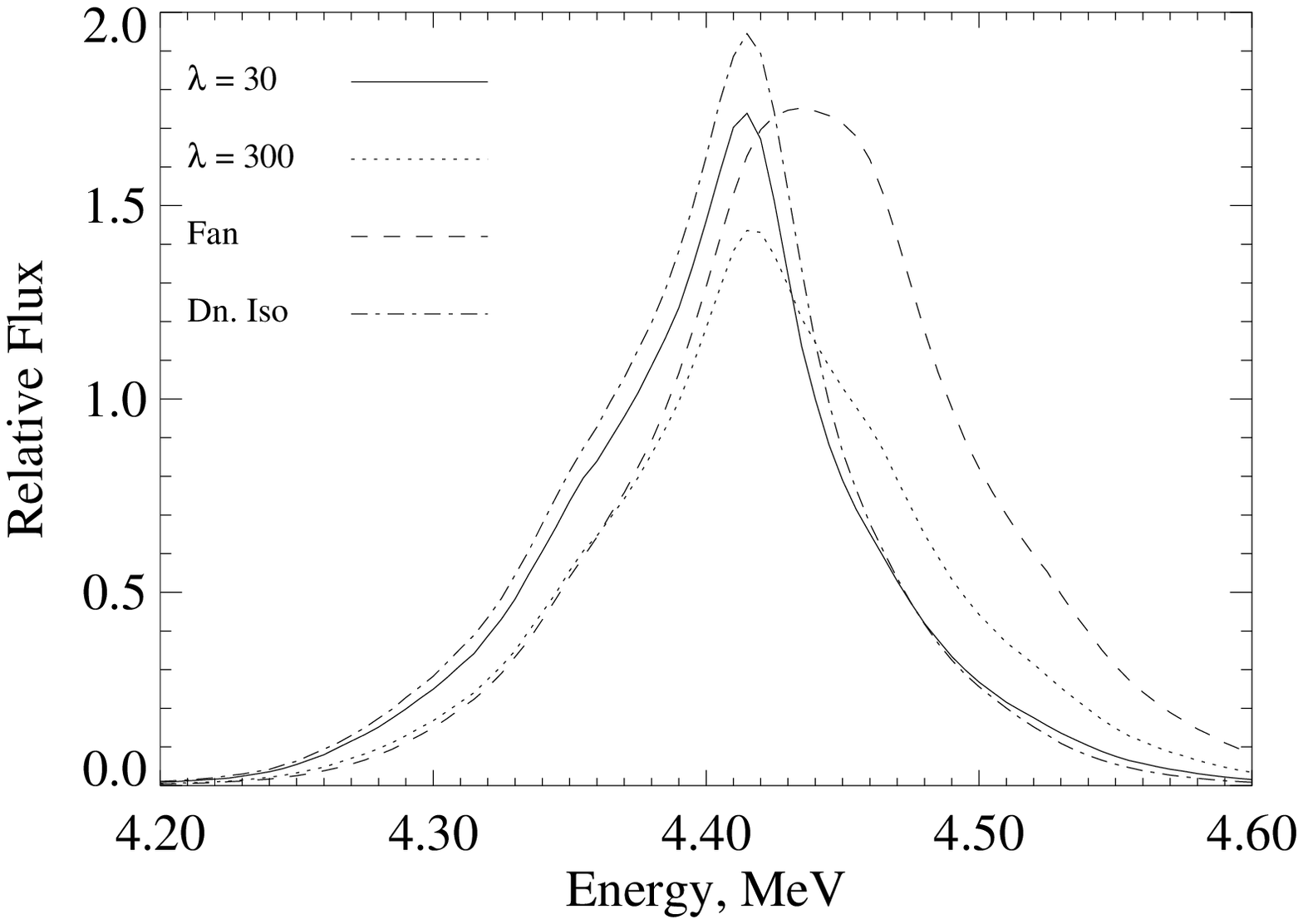}
\caption{Representative particle distributions (left panel) used to calculate $^{12}$C line shapes (right panel) for a heliocentric angle of 30$^{\circ}$, $\alpha$/p = 0.5, and power law spectral index 4.  $\theta$ is the angle from the outward radial direction.} \label{fig10}
\end{figure}
\clearpage

\clearpage

\begin{deluxetable}{rrrrrrrr}
\tabletypesize{\scriptsize}
\tablecaption{Probabilities for Carbon Line Fits.\label{tbl-1}}
\tablewidth{0pt}
\tablehead{
\colhead{Angle} & \colhead{Gaussian}   & \colhead{Dn. Beam}   &
\colhead{Fan Beam} &
\colhead{Iso.}  & \colhead{Dn. Iso.} & \colhead{$\lambda = 30$}  
& \colhead{$\lambda = 300$}}
\startdata
 5$^{\circ}$&0.35 &0.12 &0.19 &0.22 & 0.20 (0.23)& 0.21 & 0.22\\
 30$^{\circ}$&0.32 &0.04 &0.007 &0.007 &0.31 (0.29) & 0.33 & 0.15 \\
 43$^{\circ}$& 0.18 &$0.002$ &0.03  &0.03 & 0.15 (0.24) & 0.18 & 0.20\\
 74$^{\circ}$& 0.17 &$0.01$ &0.22 & 0.22&0.16 (0.20)& 0.17 & 0.23 \\
 86$^{\circ}$& 0.77 &0.40 &0.76 & 0.71&0.70 (0.71)  & 0.70 & 0.74 \\

 \enddata
\tablecomments{Probabilities that the value of $\chi ^2$ for a random distribution of data about a mean exceeds the values obtained in the line fit for the same number of degrees of freedom.  The probabilities listed in the parentheses in the 6$^{th}$ column were derived for a combination of downward isotropic and upward isotropic distributions in the ratio of 5:1.}

\end{deluxetable}

\end{document}